*Methods to study splicing from high-throughput RNA Sequencing data*

Gael P. Alamancos[1], Eneritz Agirre[1], Eduardo Eyras[1,2,*]

[1]Universitat Pompeu Fabra, Dr Aiguader 88, E08003 Barcelona, Spain
[2]Catalan Institution for Research and Advanced Studies (ICREA), Passeig Lluís Companys 23, E08010 Barcelona, Spain
[*]eduardo.eyras@upf.edu

1. Introduction

The development of novel high-throughput sequencing (HTS) methods for RNA (RNA-seq) has facilitated the discovery of many novel transcribed regions and splicing isoforms (Djebali et al. 2012) and has provided evidence that a large fraction of the transcribed RNA in human cells undergo alternative splicing (Wang et al. 2008, Pan et al. 2008). RNA-Seq thus represents a very powerful tool to study alternative splicing under multiple conditions at unprecedented depth. However, the large datasets produced and the complexity of the information to be analyzed has turned this into a challenging task. In the last few years, a plethora of tools have been developed (Figure 1), allowing researchers to process RNA-Seq data to study the expression of isoforms and splicing events, and their relative changes under different conditions. In this review, we provide an overview of the methods available to study alternative splicing from short RNA-Seq data. We group the methods according to the different questions they address:

1) Assignment of the sequencing reads to their likely gene of origin. This is addressed by methods that map reads to the genome and/or to the available gene annotations (Table1)
2) Quantification of events and isoforms. Either after reconstructing transcripts or using an annotation, these methods estimate the expression level or the relative usage of isoforms and/or events (Tables 2, 3 and 5)
3) Recovering the sequence of splicing events and isoforms. This is addressed by transcript reconstruction and *de novo* assembly methods (Tables 4 and 6).
4) Providing an isoform or event view of differential splicing or expression. These include methods that compare relative event/isoform abundance or isoform expression across two or more conditions (Tables 7 and 8).
5) Visualizing splicing regulation. Various tools facilitate the visualization of the RNA-Seq data in the context of alternative splicing (Table 9).

In this review, we use transcript or isoform to refer to a distinct RNA molecule transcribed from a gene locus. We use gene to refer to the set of isoforms transcribed from the same genomic region and the same strand, sharing some exonic sequence; and a gene locus refers to this genomic region. A splicing event refers to the exonic region of a gene that shows variability across two or more of its isoforms. Splicing events generally include exon skipping (or cassette exon), alternative 5' and 3' splice-sites, mutually exclusive exons, retained introns, alternative first exons and alternative last exons (see for example (Chen 2011)), although other events may occur as a combination of two or more of these ones. In this review, we do not enter into the details of the specific mathematical models behind each method. For a comparative analysis of the mathematical models used by many of these methods see (Pachter 2011). Our aim is rather to provide an overview that could serve as an entry point for users who need to decide on a suitable tool for a specific analysis. We also attempt to propose a classification of the tools according to the operations they do, to facilitate the comparison and choice of methods.

## 2. Spliced-mapping short reads

Event and Isoform quantification are very much dependent on the correct assignment of RNA-Seq reads to the molecule of origin. Accordingly, we will start by reviewing some of the read mappers that are splice-site aware, and therefore, can be used to detect exon-intron boundaries and connections between exons. This alignment problem has been addressed in the past by tools that combine fast heuristics for sequence matching with a model for splice-sites, for example Exonerate (Slater et al. 2005), BLAT (Kent 2002), or GMAP (Wu et al. 2005). These methods, however, are not competitive enough to map all reads from a sequencing run in a reasonable time. In the last few years, a myriad of methods have been developed for mapping short reads to a reference genome (Fonseca et al. 2012). Those that are splice-site aware and incorporate intron-like gaps are generally called spliced-mappers, split-mappers, or spliced aligners. Their main challenge is that reads must be split into shorter pieces, which may be harder to map unambiguously; and although introns are marked by splice-site signals, these occur frequently by chance in the genome.

Spliced-mappers have been classified previously into two main classes (Garber et al. 2011), *exon-first* and *seed-and-extend* (Table 1). *Exon-first* methods map reads first to the genome using an unspliced approach to find read-clusters; unmapped reads are then used to find connections between these read-clusters. These methods include TopHat (Trapnell et al. 2009), MapSplice (Wang et al. 2010a), SpliceMap (Au et al. 2010), HMMsplicer (Dimon et al. 2010), SOAPsplice (Huang et al. 2011), PASSion (Zhang et al. 2012), TrueSight (Li et al. 2012b), and GEM (Marco-Sola et al. 2012). *Seed-and-extend* methods generally start by mapping part of the reads as *k*-mers or substrings; candidate matches are then extended using different algorithms and potential splice-sites are located. These methods include MapNext (Bao et al. 2009), PALMapper (Jean et al. 2010), SplitSeek (Ameur et al. 2010), GSNAP (Wu et al. 2010), Supersplat (Bryant et al. 2010), SeqSaw (Wang et al. 2011), STAR (Dobin et al. 2012). A generalization of seed-and-extend methods is represented by the multi-seed methods, like CRAC (Philippe et al. 2013), OLego (Wu et al. 2013), and Subread (Liao et al. 2013), which consider multiple subreads within each read. Similarly, ABMapper (Lou et al. 2011) consider multiple read-splits for mapping. Some methods actually use a hybrid strategy, following an exon-first approach for unspliced reads, and then using seed-and-extend approach for spliced reads, like MapSplice, SpliceMap, HMMSplicer, TrueSight, GEM, and PALMapper; the latter being a combination of GenomeMapper (Schneeberger et al. 2009) and QPalma (De Bona et al. 2008) for spliced reads. *Exon-first* methods depend strongly on sufficient coverage on potential exons to incorporate spliced reads, but are generally faster than *seed-and-extend* methods. On the other hand, *seed-and-extend* methods tend to be less dependent on recovering exon-like read-clusters and may recover more novel splice-sites. However, the storage of *k*-mers for long reads requires sufficient computer memory for large *k*, and the mapping has limited accuracy for small *k* (Huang et al. 2011).

There is also a different class of tools, which use the annotation and/or some heuristics to map reads. These include X-Mate (Wood et al. 2011), SAMMate (Xu et al. 2011), IsoformEx (Kim et al. 2011), RUM (Grant et al. 2012), MapAI (Labaj et al. 2012), RNASEQR (Chen et al. 2012b), SpliceSeq (Ryan et al. 2012), OSA (Hu et al. 2012) and PASTA (Tang et al. 2013). RNASEQR and RUM use Bowtie (Langmead et al. 2009) to map reads to the transcriptome and genome; and then identify novel junctions from the unmapped reads using BLAT (Kent 2000). Similarly, SAMMate and IsoformEx use Bowtie to locate reads in exons and junctions, whereas SpliceSeq uses Bowtie to map reads to a graph representation of the annotation; X-Mate uses its own heuristics to trim and map reads recursively to locate reads on exons and junctions. On the other hand, PASTA does not use any gene annotation; it uses Bowtie and a splice-site model to locate read fragments on exon junctions. Among these methods, SAMMate, IsoformEx, RUM, and SpliceSeq also provide some level of quantification for exons, events, or isoforms (Table 2) (Figure 1), which makes them convenient as a pipeline tool. OSA is actually a seed-and-extend mapping method but relies on an annotation. OSA avoids splitting reads into subreads, which helps improving speed; and like other annotation-guided methods, also split-maps reads that are not located in the provided annotation using the seed-and-extend approach. Finally, unlike the other methods, MapAI and ContextMap use reads already mapped to a reference genome. MapAI uses reads mapped to a transcriptome to assign them to their genomic positions, whereas ContextMap refines the genome mappings using the read context, extending to all reads the context approach used by methods like MapSplice or GEM for spliced reads. In the newest version, ContextMap can also be used as a standalone read-mapping tool. Annotation-guided mapping methods are possibly the best option to accurately assign reads to gene annotations, whereas *de novo* mapping tools are convenient for finding new splicing junctions.

Besides the differences in the mapping procedure, *de novo* mapping tools detect splice-sites using a variety of approaches, which may determine the reliability of the splice-sites detected and the possibility of obtaining novel ones. Most tools search for an exact match of the flanking intronic dinucleotides to the canonical splice-sites GT-AG, GC-AG, AT-AC (Table 1). Tools like MapNext and Tophat use a two-step approach, first mapping to the known junctions and then locating novel ones with GT-AG dinucleotides, whereas tools like MapSplice, Supersplat, SpliceMap, and HMMSplicer use a gapped-alignment approach that allows the detection of junctions regardless of the exon coverage. HMMSplice, QPalma, PASTA, and OLego use a more complex representation for splice-sites. HMMSplice is based on a hidden Markov model, QPalma on a Support Vector Machine, PASTA on a logistic regression, and OLego on the combined logistic modeling of sequence bias and intron-size; all of which are trained on known splice-sites. In contrast, MapSplice, SeqSaw, STAR, SplitSeek, and CRAC can do an unbiased search of splice-junctions, not necessarily looking for the splice-site motif and generally using support from multiple reads; hence, they can potentially recover noncanonical splice-sites. Annotation-guided methods will accurately assign reads to known splice-sites, but will miss novel ones, unless they use some heuristics for novel junctions like RUM and RNASEQR. Mapping methods like STAR, GEM, MapNext, and TopHat accept annotation as optional input, which will guide the initial mapping of reads. Other parameters may be important too, like the search range of intron lengths. Most models impose restrictions in the minimum and maximum intron lengths, but methods like MapSplice does not impose any restriction and OSA has a specific search for novel exons using distal fragments. The decision of which tool to use depends very much on whether the aim is either to assign reads to known annotations or to find novel splice-sites.

## 3. Definition and quantification of events and isoforms

First reports using RNA-Seq to quantify splicing followed an approach analogous to splicing junction arrays (Clark et al. 2002). They were based on the analysis of junctions built from known gene annotations (Cloonan et al. 2009, Pan et al. 2008, Sultan et al. 2008, Wang et al. 2008, Tang et al. 2009). In these and later methods, reads aligning to candidate alternative exons and its junctions are considered as inclusion reads, whereas reads mapping to flanking exons and to junctions skipping the candidate alternative exon are considered as skipping or exclusion reads. These reads are then used to provide an estimate of the relative inclusion of the regulated exon (Chen 2012), generally called inclusion level. This approach has shown a reasonable agreement with microarrays and can be modified to include exon-body reads and variable exon lengths (Wang et al. 2008, Chen et al. 2012).

An alternative measure, "percent spliced in" (PSI), was defined as the number of isoforms that include the exon over the total isoforms (Venables et al. 2008), or equivalently, as the fraction of mRNAs that represent the inclusion isoform (Katz et al. 2010). If the PSI value is calculated for a particular splicing event, it can be considered equivalent to the inclusion level. Isoform quantification can be expressed in terms of either a global measure of expression (Glaus et al. 2012), which may provide a global ranking comparable across genes in one sample, or in terms of a relative measure of expression, which is normalized per gene locus and comparable across conditions. The global measure is generally given in terms of RPKM or FPKM (Reads or Fragments Per Kilobase of transcript sequence per Millions mapped reads); and the relative measure is given in terms of a PSI value or a similar value.

Besides the original approaches (Cloonan et al. 2009, Pan et al. 2008, Sultan et al. 2008, Wang et al. 2008, Tang et al. 2009), various tools have been developed recently to quantify events and isoforms. These range from simply quantifying the inclusion of events to the reconstruction and quantification of novel isoforms. Some of the tools that reconstruct isoforms also estimate their quantification, and some tools may quantify either known isoforms or novel ones, or both simultaneously. Accordingly, we classify the methods depending on whether they use annotation or not, and on the type of input and output:
1) Event/isoform quantification using known (genome-based) gene annotations (Table 2)
2) Isoform quantification using a transcriptome annotation (Table 3).

3) *De novo* isoform reconstruction with a genome reference, either purely focused on reconstruction or also providing isoform quantification (Table 4).
4) Isoform reconstruction and quantification guided by annotation. These methods use a gene annotation as a guide and can complete the annotation with new exons, new isoforms, or even with some new gene loci (Table 5).
5) Finally, some of the *de novo* transcriptome assembly methods also quantify isoforms (Table 6)

## 3.1 Event and Isoform quantification guided by gene annotation

Various tools have been developed for event quantification from a single condition (Table 2) (Figure1): MISO (Katz et al. 2010), ALEXA-Seq (Griffith et al. 2010), SOLAS (Richard et al. 2010), RUM (Grant et al. 2012), SpliceTrap (Wu et al. 2011), MMES (Wang et al. 2010b), and SpliceSeq (Ryan et al. 2012). RUM provides quantification of genes, exons and junctions in terms of read-counts and RPKM (reads per kilobase per million mapped reads), whereas, MMES use the reads mapped to junctions and employ a statistical model to calculate exon inclusion levels and junction scores. RUM and MMES also provide the mapping step. RUM has its own heuristics (Table 1), whereas MMES maps reads to exon-exon junctions using SOAP (Li et al. 2009). Similarly, SpliceSeq maps reads to a splicing-graph to obtain exon and junction inclusion levels. MISO and ALEXA-Seq use reads on exons and junctions, whereas SOLAS uses only reads on exons. MISO provides PSI values, while ALEXA-Seq and SOLAS provide event and isoform expression levels. MISO, ALEXA-Seq, and SOLAS can also estimate isoform relative abundances and can be further used for differential splicing (Table 7).

Quantification of isoforms is more complicated than that of events, as it requires the correct assignment of reads to isoforms sharing part of their sequence. One of the first attempts to do this was Erange (Mortazavi et al. 2008), where reads mapped to the genome and known junctions were distributed in isoforms according to the coverage of the genomic context, and isoform expression was defined in terms of RPKM. However, the uncertainty in the assignment of reads shared by two or more isoforms must be appropriately modeled. Accordingly, a number of methodologies have been proposed to address this issue (Table 2): rSeq (Jiang et al. 2009), rQuant (Bohnert et al. 2009), SOLAS (Richard et al. 2010), MISO (Katz et al. 2010), ALEXA-Seq (Griffith et al. 2010), Cufflinks (Trapnell 2010), IsoInfer (Feng et al. 2010), FluxCapacitor (Montgomery et al. 2010), SAMMate (Xu et al. 2011), IsoformEx (Kim et al. 2011), SLIDE (Li et al. 2011b), DRUT (Mangul et al. 2012), iReckon (Mezlini et al. 2013), IQSeq (Du et al. 2012), RABT (Roberts et al. 2012), and Casper (Rossell et al. 2012). Isoform quantification is generally given in terms of RPKM, FPKM or some equivalent *isoform expression level* value; or in terms of a *relative expression* value like PSI or equivalent.

SAMMate and IsoformEx use the reads mapped to exons and junctions by their own methods to quantify gene and isoform expression in terms of RPKM values. SAMMate incorporates two quantification methods, one that is not sensitive to coverage, so it can be used on early sequencing platforms (Deng et al. 2011) and a recent one that is aimed for deeper coverage and uses a filtering of non-expressed transcripts (Nguyen et al. 2013), SOLAS and rSeq use reads on exons to estimate isoform expression levels; whereas rQuant uses the position-wise density of mapped reads to calculate two abundance estimates: the RPKM and the estimated average read coverage for each transcript. IQSeq provides a statistical model that facilitates the incorporation of data from multiple technologies; and FluxCapacitor, unlike other methods, does not account for the mapping variability across isoforms and directly solves the constraints derived from distributing the reads in isoforms according to the splicing graph built from the read evidence.

IsoInfer, SLIDE, RABT, DRUT, and iReckon can quantify the known annotation and at the same time predict and quantify novel isoforms in known gene loci. RABT quantifies known and novel isoforms, taking into account existing gene annotations and using the same graph assembly algorithm of Cufflinks, combining the sequencing reads with reads obtained by fragmenting known transcripts. RABT is part of the Cufflinks distribution, but here we distinguish it from the original Cufflinks, which quantifies abundances of either only annotated or only novel isoforms (Trapnell et al. 2010, Roberts et al. 2012). Similar to RABT, SLIDE uses RNA-Seq data and existing gene annotation to discover novel isoforms and to estimate the abundance of known and new isoforms. Additionally, it can use other sources of evidence, like RACE, CAGE, and EST, or even the output from other isoform reconstruction algorithms. IsoInfer uses the transcript start and end sites, plus exon-intron boundaries to enumerate all possible isoforms, estimate their expression levels and then choose the subset of isoforms that best explain the observed reads, predicting novel isoforms from the existing exon data. On the other hand, iReckon can work with just transcript start and end sites or with full annotations; it models multimapped reads, intron-retention and unspliced pre-mRNAs and performs reconstruction and quantification simultaneously. DRUT uses a modified version of the IsoEM algorithm (Nicolae et al. 2011) in combination with a *de novo* reconstruction method similar to Cufflinks to complete partial existing annotations as well as to estimate isoform frequencies. Casper, similar to Cufflinks, estimates abundances of known or novel isoforms separately, but unlike other methods, uses information of the connectivity of more than two exons. Generally, known isoform quantification methods show a high level of agreement with experimental validation (Mezlini et al. 2013) and can be improved using annotation-guided methods for read mapping (Labaj et al. 2012).

3.2 Isoform quantification guided by a transcriptome

A number of methods consider reads mapped to a transcriptome for isoform quantification (Table 3); these include RSEM (Li et al. 2011c), IsoEM (Nicolae et al. 2011), NEUMA (Lee et al. 2011), BitSeq (Glaus et al. 2012), MMSEQ (Turro et al. 2011), and eXpress (Roberts et al. 2013). Although these methods depend on a transcriptome annotation, they can use a standard (non-spliced) mapper to obtain the input data. Additionally, they can work also with predicted isoforms from transcript assembly methods (Figure 1). All of them provide a measure of global isoform expression, similar to RPKM. Moreover, RSEM also calculates the fraction of transcripts represented by the isoform, equivalent to PSI. RSEM and IsoEM use both an Expectation-Maximization algorithm and model paired-end fragment size. RSEM models the mapping uncertainty to transcripts and provides confidence intervals of the abundance estimates. IsoEM uses the fragment-size information to disambiguate the assignment of reads to isoforms. BitSeq is based on a Bayesian approach, incorporates the mapping step to the transcriptome, models the nonuniformity of reads and provides an expression value per isoform. BitSeq can also be used for differential isoform expression (see below). MMSEQ also takes into account the nonuniform read distribution and deconvolutes the mapping to isoforms to estimate isoform-expression and haplotype-specific isoform-expression. The method eXpress is in fact a general tool for quantifying abundances of a set of sequences in a generic experiment and can be used with a reference genome or transcriptome. For RNA-Seq reads mapped to a transcriptome, eXpress provides isoform quantification in terms of FPKM. Finally, NEUMA is different from the other methods, as it does not use any probabilistic description and assumes uniformity of the reads along transcripts. NEUMA labels reads according to whether they are isoform or gene specific and calculates a measure of isoform quantification defined as the number of fragments per virtual kilobase per million reads (FVKM). Transcript-based methods can be generally applied to the transcripts obtained from genome annotations, so that the correspondence of transcripts to gene loci is maintained. Additionally, they can be used in combination with *de novo* transcript assembly methods (see below) to estimate isoform abundance in genomes without a reference.

## 3.3 Genome-based transcript reconstruction and quantification without annotation

These methods use the reads mapped to the genome to reconstruct isoforms *de novo*. They are generally based on previous approaches to transcript reconstruction from ESTs (Heber et al. 2002, Haas et al. 2003, Xing et al. 2004, Xing et al. 2006). As for ESTs (Nagaraj et al. 2007), accuracy is limited by the length of the input reads; hence, the use of paired-end sequencing may provide some improvements. Additionally, as RNA abundance spans a wide range of values, the correct recovery of lowly expressed isoforms requires sufficient sequencing coverage. Although these methods work independently of the mapping procedure, they strongly rely on the accuracy of the spliced-mapper.

Purely reconstruction methods, without isoform quantification, include G-Mo.R-Se (Denoeud et al. 2008) and assemblySAM (Zhao et al. 2011a). Methods that reconstruct isoforms as well as estimate their abundances include Cufflinks (Trapnell et al. 2010), TAU (Filichkin et al. 2010), Scripture (Guttman et al. 2010), IsoLasso (Li et al. 2011a), CEM (Li et al. 2012a), NSMAP (Xia et al. 2011), Montebello (Hiller et al. 2012), and Casper (Rossell et al. 2012). G-Mor.R-Se, Scripture and TAU proceed in a similar way by first obtaining candidate exons from read-clusters and then connecting them using reads spanning exon-exon junctions. Subsequently, all possible isoforms from the graph of connected exons are computed. As they explore all possible connections between potential exons, they ensure a high sensitivity but at the cost of a high false positive rate. In contrast, Cufflinks first connects predicted exons trying to identify the minimum number of possible isoforms using a graph generated from the reads; expression levels are then calculated using a statistical model (Jiang et al. 2009). IsoLasso also tries to obtain the minimal set of isoforms from predicted exons, but maximizing the number of reads included in each isoform. CEM model takes into account positional, sequencing and mappability biases of the RNA-Seq. Casper follows a heuristics similar to Cufflinks but exploiting the reads that connect more than 2 exon. Some of these methods use paired-end reads and model the insert-size distribution, which both improve the reconstruction accuracy (Salzman et al. 2011). NSMAP, IsoLasso, and Montebello perform identification of the exonic structures and estimation of the isoform expression levels simultaneously in a single probabilistic model; iReckon does so too, but was not included in this section as it needs at least the transcript start and end positions. The rest of methods perform reconstruction and quantification independently.

Although a large overlap among methods has been reported (Li et al. 2010), there are still many predictions unique to each method. Interestingly, given a fixed number of sequenced bases, sequencing longer reads does not seem to lead to more accurate quantifications (Li et al. 2011c, Nicolae et al. 2011), although exonic structures may be better predicted (Li et al. 2012a). These *de novo* reconstruction and quantification methods seem a good option for finding novel isoforms (Guttman et al. 2010), alternatively spliced genes in a genome with partial annotation (Denoeud et al. 2008) and for quantifying isoforms under various conditions (Trapnell et al. 2010). However, they depend much on coverage. Accordingly, if the aim is to obtain the expression of known isoforms, gene-based methods may be a better choice. Alternatively, for protein-coding gene finding there are other options available, as discussed next.

## 3.4 Evidence-based alternatively spliced gene prediction

The methods described above are mainly focused on isoform quantification using available annotation or on the *de novo* reconstruction and quantification of isoforms, using reads mapped to the genome. Quantification methods based solely on gene annotations could miss many novel genes and isoforms, whereas *de novo* approaches not using annotations may produce many false positives. Combined approaches that discover novel isoforms in known and new loci and, at the same time, quantify them, could help improving the gene annotation. Some of the annotation-based quantification methods can also reconstruct and quantify new isoforms in known gene loci (Table 5): RABT (Roberts et al. 2012), IsoInfer (Feng et al. 2012), SLIDE (Li et al. 2011b), iReckon (Mezlini et al. 2013), and DRUT (Mangul et al. 2012). Some of these methods can work with partial evidence, like iReckon. However, they do not predict new isoforms in new gene loci. To this end, a number of methods can use RNA-Seq and other sources of evidence to predict the exon-intron structures of isoforms, or even to predict full protein-coding gene structures. These methods include (Table 5) SpliceGrapher (Rogers et al. 2012), TAU (Filichkin et al. 2010), mGene (Behr et al. 2010), and the method described in (Seok et al. 2012a). The method mGene is an SVM-based gene predictor (Sonnenburg et al. 2007) that first reconstructs a high-quality gene set, which then uses to train a gene model that is applied using RNA-Seq data in addition to the previously determined genomic signal predictors. In contrast, SpliceGrapher and TAU incorporate into the same graph model information from ESTs and RNA-Seq reads to complete known gene annotations and produce novel variants. ExonMap/JunctionWalk, proposed in (Seok et al. 2012a), combine SpliceMap (Au et al. 2010) alignments with known annotations to complete known isoforms and obtain novel ones without quantification (Figure 1).

Some of these methodologies are reminiscent of the evidence-based gene prediction methods. These are generally based on probabilistic models of protein-coding genes, which can incorporate external spliced evidence like ESTs and cDNAs into the model to guide the prediction of the exon-intron structure, and some of which can predict multiple isoforms in a gene locus. Accordingly, evidence-based gene prediction methods could still be useful for splicing analysis from RNA-Seq. In particular, Augustus (Stanke et al. 2006a) is an evidence-based protein-coding gene prediction method, capable of finding multiple isoforms per gene, which has been shown to be highly accurate using a blind test set (Stanke et al. 2006b, Guigó et al. 2006). Other evidence-based prediction methods include (Table 5) GAZE (Howe et al. 2002), JigSaw (Allen et al. 2005), EVM (Haas et al. 2008), and Evigan (Liu et al. 2008). Although these four methods do not explicitly model alternative isoforms, they can still produce multiple transcripts in a locus.

Evidence-based gene prediction methods can take as input transcripts reconstructed by other methods and generate protein-coding isoforms. They do not provide a quantification of isoforms, but in combination with quantification methods (Tables 2 and 3) they could be a powerful approach to annotate and quantify alternatively spliced protein-coding genes from newly sequenced genomes using RNA-Seq data.

### 3.5 *De novo* transcript assembly

*De novo* transcript assemblers put together reads into transcriptional units without mapping the reads to a genome reference, similar to building Unigene clusters from ESTs prior to having a genome reference (Pontius et al. 2003). A transcriptional unit can be defined as the set of RNA sequences that are transcribed from the same genome locus and share some sequence, i.e., the set of RNA isoforms from the same gene. This is generally represented as a sequence-based graph, where paths along the graph potentially resolve the different isoforms. Methods for transcript assembly include (Table 6) TransAbyss (Robertson et al. 2010), Rnnotator (Martin et al. 2010), STM (Surget-Groba et al. 2010), Trinity (Grabherr et al. 2011), SOAPdenovo-trans (Li et al. 2009), KisSplice (Sacomoto et al. 2012), and OASES (Schulz et al. 2012). Although KisSplice focuses on recovering alternative splicing events, we include it here as it follows a similar approach to the other methods. See (Zhao et al. 2011b) for a recent comparison between some of these methods.

The main challenge of these methods is not only to distinguish sequence errors from polymorphisms but also to distinguish close paralogues from alternative isoforms, which requires correctly capturing the exonic variability. All these methods are based on a graph built from k-mer overlaps between read sequences. The choice of *k*-mer length affects the assembly, being more sensitive at low values of *k*, and more specific at high values. Accordingly, some use a variable k-mer approach. Isoforms are recovered as paths through the graph with sufficient read coverage. Not all methods can provide multiple isoforms from the same gene (Table 6).

Genome-independent methods are useful when there is no genome reference sequence available, and could also be valuable when the RNA is expected to contain much variation, like in a cancer cell with many copy number alterations, mutations and genome rearrangements compared to the reference genome. *De novo* assembly methods tend to be more sensitive to sequencing errors and low coverage, and generally require more computational resources, although full parallelization of the graph algorithms can alleviate this issue (Jackson et al. 2009). Some of the methods also consider the comparison to reference sets of DNA or protein sequences (Surget-Groba et al. 2010). In fact, mapping assembled transcripts to a reference genome, even from a related species, seems to improve accuracy in transcript quantification (Vijay et al. 2013). KisSplice is explicitly designed to obtain and quantify *de novo* alternative splicing events, which may potentially be coupled with other methods to study differential splicing. On the other hand, OASES, TransAbyss, Trinity, and SOAPdenovo-trans can produce multiple isoforms, but only TransAbyss and Trinity perform quantification. Nonetheless, multiple assembled isoforms can be quantified with transcript-based methods (Table 3) or further processed with isoform-based differential expression methods (Table 8).

## 4. Comparing splicing across samples

The comparison of events and isoforms across two or more conditions provide valuable information to understand the regulation of alternative splicing. However, it is important to distinguish differential isoform relative abundance, from differential isoform expression. Changes in relative abundance of isoforms, regardless of the expression change, indicate a splicing-related mechanism. On the other hand, there can be measurable changes in the expression of isoforms across samples, without necessarily changing the relative abundance, which possibly indicates a transcription-related mechanism. With this in mind, we can consider two types of methods, those that measure relative event or isoform usage (Table 7) and those that measure isoform-based changes in expression (Table 8).

### 4.1. Differential splicing

Most of these methods are focused on splicing events, thereby summarizing the isoform relative abundance into two possible splicing outcomes in a local region of the gene (Figure 1). They use a predetermined set of splicing events, generally calculated from gene annotations and additional EST and cDNA data; hence, they are suitable for studying splicing variation in well-annotated genomes. They all consider exon-skipping events (cassette exons), and some also include alternative 5' and 3' splice-sites, mutually exclusive exons and retained introns; and in very few cases, multiple-cassette exons, alternative first exons and alternative last exons (Katz et al. 2010). Potential novel events are sometimes built by considering hypothetical exon-exon junctions from the annotation (Shen et al. 2011).

Methods that calculate differential relative abundance of events or exons under at least two conditions include (Table 7) SOLAS (Richard et al. 2010), ALEXA-Seq (Griffith et al. 2010), MISO (Katz et al. 2010), GPSeq (Srivastava et al. 2010), MATS (Shen et al. 2011), JuncBase (Brooks et al. 2011), DEXSeq (Anders et al. 2012),

DSGSeq (Wang et al. 2013), SpliceSeq (Ryan et al. 2012), JETTA (Seok et al. 2012b), rDiff (Drewe et al. 2013, Stegle et al. 2010), FDM (Singh et al. 2012), DiffSplice (Hu et al. 2013), SplicingCompass (Aschoff et al. 2013), and the methods from (Kakaradov et al. 2012). ALEXA-Seq estimates inclusion levels on a set of pre-calculated events using only unambiguous reads, i.e., reads that map to one unique event, and calculates various measures of differential expression, including the splicing index, i.e., a measure of change in expression of an event between two conditions relative to the change in expression of the entire gene locus between the same two conditions. On the other hand, SOLAS uses single-reads and only takes into account those mapping within exons, disregarding reads spanning exon-exon junctions, to detect differentially spliced events between two conditions. DEXSeq, DSGSeq, and GPSeq use read counts on exons to calculate those genes with differential splicing between two conditions. They do not provide any event or isoform information and report the exon with significant change (Figure 1). MATS and MISO use both a Bayesian approach to calculate the differential inclusion of splicing events between two samples, using reads that map to exons and to the inclusion and skipping exon junctions. JuncBASE also uses reads mapped to exon junctions and uses a Fisher exact test to compare the read count in the inclusion and exclusion forms in two conditions. JETTA estimates the differential inclusion between two conditions from pre-calculated expression values for genes, exons, and junctions, which the authors obtain using SeqMap (Jiang et al. 2008) and rSeq (Jiang et al. 2009). SpliceSeq calculates read coverage along genes, exons, and junctions for each sample, which are then compared to identify significant changes in splicing across samples. SpliceSeq also includes the evaluation of the impact of alternative splicing on protein products and a visualization of the events (see below). Besides all these methods, various methods were proposed in (Kakaradov et al. 2012) based on reads over exon junctions to find robust estimates of PSI, taking into account the positional bias of reads relative to the junction.

Some of these methods can also measure the change in the relative abundance of isoforms (Figure 1): MISO can measure changes in isoform relative abundances from previously calculated isoform PSI values; ALEXA-Seq uses the events that are differentially expressed to infer isoform abundance differences between two conditions. Finally, rDiff, FDM, and DiffSplice are methods that work with a more general definition of event and that can operate without an annotation. FDM and DiffSplice are graph-based methods and both identify regions of differential abundance of transcripts between two samples using the variability of reads that define a splicing graph. Similarly, rDiff uses a Maximum Mean Discrepancy test (Borgwardt et al. 2006) to estimate regions that have a significant distance between the read distributions in the two conditions. Alternatively, rDiff can work with an annotation; it considers reads in exonic regions that are not in all isoforms and groups those regions according to whether they occur in the same set of isoforms. Finally, SplicingCompass uses a geometric approach to detect differentially spliced genes and quantifies relative exon usage. In summary, these methods test whether events, isoforms, or genic regions, change their relative abundances between two or more conditions, and so directly address the question of differential splicing.

When comparing two or more conditions, biological variability becomes an important issue, which has been shown to be relevant for studying expression (Hansen et. 2011) and splicing (Anders et al. 2012) from RNA-Seq data. However, not all methods take this into account. From the methods described here, DEXSeq, DSGSeq,

GPSeq, DiffSplice, FDM, rDiff, and a newer version of MATS accept multiple replicates and model biological variability in different ways. In contrast, the initial methods for calculating splicing changes from RNA-Seq data (Pan et al. 2008, Sultan et al. 2008, Wang et al. 2008), as well as MISO, ALEXA-Seq, JETTA, SpliceSeq, SOLAS and SplicingCompass do not work with multiple replicates. On the other hand, JuncBASE can work with replicated data but does not seem to model variability. As the cost of sequencing continue to decrease, it will be more common to include replicates in the differential splicing analysis, which will prove relevant to discern actual regulatory changes from biological variability.

## 4.3. Isoform-based differential expression

Current methods to study differential splicing at the event level show a high validation rate (Pan et al. 2008, Shen et al. 2012). However, their agreement with microarray-based methods is not as high as one may expect (Pan et al. 2008). This limitation could be due to the simplification of considering only events, rather than full RNA isoforms. An improvement in this direction would be to quantify changes in isoform expression. A possible approach is to combine methods that quantify isoforms with methods for differential gene expression. However, as previously pointed out (Pachter 2011, Singh et al. 2012, Trapnell et al. 2012), this may be problematic, since tools for differential gene expression analysis do not generally take into account the uncertainty of mapping reads to isoforms. We will not discuss here the many methods that have been proposed to study differential gene expression analysis from RNA-Seq data; for a recent review see (Oshlack et al. 2010, Pachter 2011).

A number of methods have been proposed to detect expression changes at the isoform level (Table 8): BASIS (Zheng et al. 2009), BitSeq (Glaus et al. 2012), Cuffdiff2 (Trapnell et al. 2012), and EBSeq (Leng et al. 2013). Cuffdiff2, BitSeq, and EBSeq take into account the read-mapping uncertainty, accept multiple replicates and model biological variability. BASIS does not accept replicates, but it models variability along genes. Cuffdiff2 and BitSeq provide quantification and differential expression of isoforms from genome-mapped and transcriptome-mapped reads, respectively. Cuffdiff2 can use reads directly mapped to the genome or can use the results from Cufflinks on two conditions after using cuffcompare (Trapnell et al. 2010) (Figure 1), which gives equivalent transcripts in both conditions. On the other hand, EBSeq relies on the isoform quantification from other methods, like RSEM or Cufflinks, and is actually included in the current release of RSEM; whereas BASIS uses coverage over exon regions that are isoform-specific to calculate differential expression of isoforms. These methods rely on an annotation, either genome-based (Cuffdiff2, BASIS, and EBSeq) or transcriptome-based (BitSeq and EBSeq). Except for Cuffdiff2, these methods do not explicitly address the question of whether the relative abundances of these isoforms change across samples (Figure 1). Accordingly, if there is an increase of transcription but the relative abundance of isoforms remain constant, they can detect changes in isoform expression, even though there might not be an actual change in splicing. On the other hand, if there are changes in the relative abundance of isoforms, they may possibly detect expression changes, but they will not provide information about the change of the relative abundances, and therefore do not directly address the question of differential splicing.

## 5. Visualizing Alternative Splicing

Being able to visualize the complexity of alternative splicing is an important aspect of the analysis. In the past, there have been multiple efforts to store and visualize alternative isoforms from ESTs and cDNAs (Bhasi et al. 2009). Visualization for RNA-Seq requires specialized tools that can efficiently process large amount of data from multiple samples. This has triggered the development of specialized tools to visualize alternative isoforms and events from RNA-Seq data (Table 9). Perhaps the simplest way to visualize isoforms and events is to generate track files for a genome browser. For instance, RSEM produces WIG files that can be viewed as tracks in the UCSC browser (Karolchik et al 2012). Similarly, SpliceGrapher and DiffSplice produce files in GFF-like formats (http://gmod.org/wiki/GFF), which can be uploaded into visualization tools like GBrowse (Donlin 2009) or Apollo (Lee et al. 2009). On the other hand, SpliceGrapher and Alexa-Seq have their own visualization utilities. Other tools have been developed independently from the analysis method. For instance, the Sashimi plot toolkit to visualize isoforms and events and their relative coverage was used with MISO but can be used with the results from other tools (Table 8). Similarly, the browser Savant (Fiume et al. 2010) has been used in conjunction with iReckon, but can be used independently for multiple HTS data formats. Finally, SpliceSeq and SplicingViewer (Liu et al. 2012) are stand-alone tools that, besides mapping reads and quantifying events and differential splicing, provides also visualization of results.

## 6. Conclusions and Outlook

The rapid development of short-read RNA sequencing technologies has triggered the development of new methods for data analysis. In this review, we have tried to provide an overview of methods applicable to the study of alternative splicing. These provide a way to detect and quantify exon-exon junctions, transcript isoforms, and differential splicing. Despite the many tools available, not all are necessarily applicable to every purpose. For instance, for genomes with good annotation coverage, like human, the expression of known isoforms and possibly their changes under several conditions might be more accurately assessed using annotation-guided methods. Similarly, if sufficient annotation is available, there are also hybrid methods that can quantify known isoforms and predict novel ones simultaneously. For newly sequenced genomes, there are effective methods to perform *de novo* reconstruction and quantification of isoforms. However, if one is specifically interested in protein-coding genes, there are also evidence-based gene prediction methods available, which can be quite effective for isoform prediction.

One can identify some open questions and areas of improvement. For instance, not all of the *de novo* transcript assembly methods describe multiple isoforms per gene and only few actually quantify them. These are still two hard problems to solve, as incompleteness or absence of transcriptomes can lead to many reconstruction and quantification errors (Pyrkosz et al. 2013). There are different approaches to improve these questions, either by a combination of methods and homology searches (Birzele et al. 2010), or by using error correction of sequencing reads before assembly (MacManes et al. 2013). These tools are of great relevance for non-model organisms and we will probably see substantial improvements in the near future. Accurate reconstruction and quantification of isoforms is crucial for downstream analysis and in particular, for differential analysis of isoform abundances. Methods to estimate differential splicing at the event level seem to provide accurate measures as shown by experimental validation. However, differential expression at the isoform level is still an active area of development.

Extending *de novo* transcriptome assembly methods to calculate differential expression of isoforms between two or more conditions could facilitate the analysis of isoform expression for non-model organisms. Although this may be done currently with a combination of methods, a tool that integrates all these could provide a powerful approach to study expression and splicing in tumor samples, where multiple genome rearrangements and copy number alterations are expected to have occurred. On a different direction, considering that a reference genome sequence does not represent all DNA that can be possibly transcribed in a cell, unmapped RNA reads may come from functional RNAs not represented in the genome annotation. Tools that map reads to a genome reference and simultaneously attempt to perform transcript assembly will be also quite useful to perform systematic analyzes of RNA in cancer samples as well as in genomes that are partly assembled.

Besides the technical improvements, there is probably also a need to improve the comparison and evaluation of current methods. Transcript reconstruction methods should be evaluated using manual gene annotation sets, as proposed previously for gene prediction methods (Guigó et al. 2006) and currently by RGASP for RNA-Seq based methods (http://www.gencodegenes.org/rgasp). Additionally, these comparisons should use measures that take into account alternative splicing (Eyras et al. 2004, Guigó et al. 2006). Similarly, there is the need to develop an experimental gold standard dataset for isoform quantification and differential isoform expression (Lovén et al. 2012).

As a final question, we may ask for how long some of these methods will be needed. There are new technologies for single-molecule sequencing that soon will be used to probe the transcriptome. This may preclude the need to perform reconstruction of isoforms. Nonetheless, short-read RNA-Seq may still be necessary for efficient quantification. On the other hand, single-molecule sequencing technologies will open up a whole new set of problems, like that of reconciling new cell-specific RNA sequences with the information available for the genome sequence and its annotation. In fact, we will be in the position to quantify multiple transcriptomes and to revisit previous studies of differential splicing and expression in cancer, as the DNA and transcription complexity of the tumor cell is fully revealed.

With this review, we have aimed to provide an overview of the different tools to study different aspects of alternative splicing from RNA-Seq data, organized such that it is useful for the end user to navigate through the list of methods. All of them have their advantages and disadvantages, but are certainly useful to answer specific questions. We also hope that this review makes it easier to identify the tools that are still missing in order to improve the study of splicing with RNA-Seq.

## Acknowledgements

We thank Y. Xing, K. Hertel, J.R. González, M. Kreitzman, and P. Drewe for comments and suggestions. This work was supported by the Spanish Ministry of Science with grants BIO2011-23920 and CSD2009-00080 and by Sandra Ibarra Foundation for Cancer with grant FSI 2011-035.

Wu J, Anczuków O, Krainer AR, Zhang MQ, Zhang C. (2013). OLego: fast and sensitive mapping of spliced mRNA-Seq reads using small seeds. Nucl. Acids Res. doi:10.1093/nar/gkt216

Xia Z, Wen J, Chang CC, Zhou X. (2011). NSMAP: a method for spliced isoforms identification and quantification from RNA-Seq. BMC Bioinformatics 12:162.

Xing Y, Resch A, Lee C. (2004). The multiassembly problem: reconstructing multiple transcript isoforms from EST fragment mixtures. Genome Res. 14(3):426-41.

Xing Y, Yu T, Wu YN, Roy M, Kim J, Lee C. (2006). An expectation-maximization algorithm for probabilistic reconstructions of full-length isoforms from splice graphs. Nucleic Acids Res. 34(10):3150-60.

Xu G, Deng N, Zhao Z, Judeh T, Flemington E, Zhu D. (2011). SAMMate: a GUI tool for processing short read alignments in SAM/BAM format. Source Code Biol Med. 6(1):2.

Zhang Y, Lameijer EW, 't Hoen PA, Ning Z, Slagboom PE, Ye K. (2012). PASSion: a pattern growth algorithm-based pipeline for splice junction detection in paired-end RNA-Seq data. Bioinformatics 28(4):479-86.

Zhao Z, Nguyen T, Deng N, Johnson K and Zhu D. (2011a). SPATA: a seeding and patching algorithm for de novo transcriptome assembly. 2011 IEEE International Conference on Bioinformatics and Biomedicine Workshop (IEEE BIBMW'11) pp. 26-33.

Zhao QY, Wang Y, Kong YM, Luo D, Li X, Hao P. (2011b). Optimizing de novo transcriptome assembly from short-read RNA-Seq data: a comparative study. BMC Bioinformatics 12 Suppl 14:S2.

Zheng S, Chen L. (2009). A hierarchical Bayesian model for comparing transcriptomes at the individual transcript isoform level. Nucleic Acids Res. 37(10):e75.

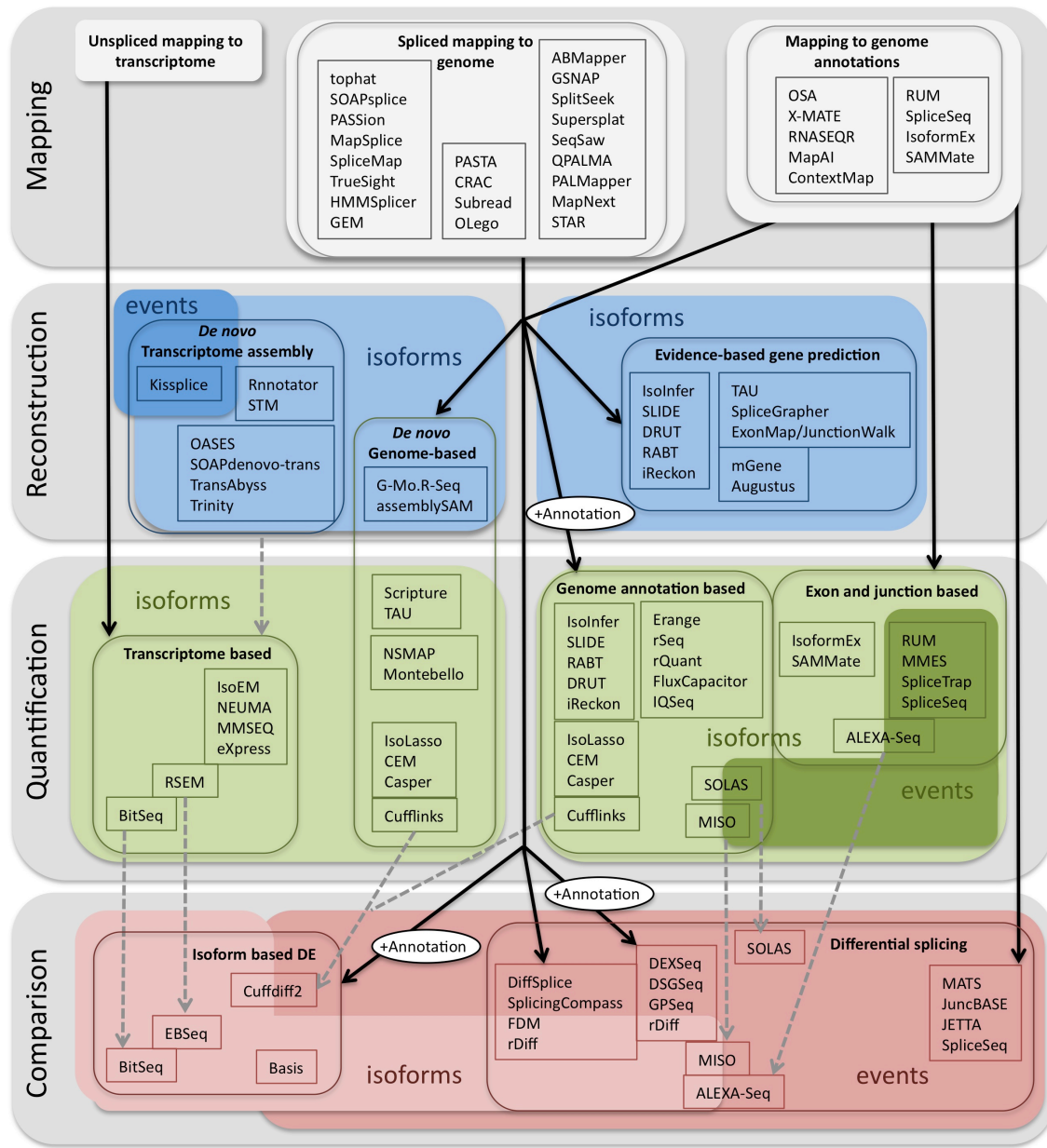

**Figure 1. Graphical representation of methods to study splicing from RNA-Seq.** Methods are divided according to whether they perform Mapping, Reconstruction of events/isoforms, Quantification of events/isoforms and whether they can perform a Comparison between two or more conditions of event/isoform relative abundances, or of isoform expression. We only list the Mapping methods that are spliced-mappers or the ones that use some heuristics to map to known exons and junctions. Methods for Reconstruction (blue), Quantification (green) and Comparison (red) are divided according to whether they work with isoforms (lighter color) or with events (darker color). Methods that work at both levels, events and isoforms, are overlapped by the two color tones. Some methods perform reconstruction and quantification and are grouped with those that only perform reconstruction. Some mapping methods also perform quantification and are repeated in two levels. Methods that require an annotation are indicated. Quantification methods that work with or without annotation are in different groups. Solid arrows connect Mapping methods to the tools in the other three levels; since, in principle, any Mapping method producing BAM as output could be fed to methods reading BAM as input. Some methods perform Mapping and Quantification or Mapping and Differential Splicing, and are connected with a solid arrow too. We indicate with dashed gray arrows those cases when a Comparison method can use the output from a Quantification method.

**Table 1. Spliced-mappers.** This table contains mapping tools that are able to locate exon-intron boundaries. Some of the methods use annotation information for mapping (OSA, X-MATE, SAMMate, IsoformEx, RNASEQR, RUM, SpliceSeq, MapAI), some can use annotation as an option (GEM, MapNext, STAR, TopHat) and others (the rest) work directly with the genome reference. Additionally, some methods perform quantification (Table 2) (SAMMate, IsoformEx, RUM, SpliceSeq) and are included here since they provide an independent method for mapping. We also indicate whether the method can map paired-end reads, the type of splice-site model, the reference where the method is described and the URL where the software is available.

| Method | Type | Uses annotation | paired-end reads | Splice site model | Reference | Web site |
|---|---|---|---|---|---|---|
| TopHat | Exon-first | Optional | Yes | Exact match to GT/C-AG | (Trapnell et al. 2009) | http://tophat.cbcb.umd.edu/ |
| SOAPsplice | Exon-first | No | Yes | Exact match to GT-AG, GC-AG, AT-AC | (Huang et al. 2011) | http://soap.genomics.org.cn/soapsplice.html |
| PASSion | Exon-first | No | Only paired-end | Exact match to GT-AG, GC-AG, AT-AC | (Zhang et al. 2012) | https://trac.nbic.nl/passion |
| MapSplice | Exon-first. Seed-and-extend for spliced reads | No | Yes | Unbiased | (Wang et al. 2010) | http://www.netlab.uky.edu/p/bioinfo/MapSplice |
| SpliceMap | Exon-first. Seed-and-extend for spliced reads | No | Yes | Exact match to GT-AG, GC-AG, AT-AC | (Au et al. 2010) | http://www.stanford.edu/group/wonglab/SpliceMap/ |
| HMMSplicer | Exon-first. Seed-and-extend for spliced reads | No | Yes | Hidden Markov Model | (Dimon et al. 2010) | http://derisilab.ucsf.edu/index.php?software=105 |
| TrueSight | Exon-first. Seed-and-extend for spliced reads | No | Yes | Exact match to GT-AG, GC-AG, AT-AC | (Li et al. 2012b) | http://bioen-compbio.bioen.illinois.edu/TrueSight/ |
| GEM | Exon-first. Seed-and-extend for splice reads | Optional | Yes | User defined regular expression and known junctions (optional) | (Marco-Sola et al. 2012) | http://algorithms.cnag.cat/wiki/The_GEM_library |
| SplitSeek | Seed-and-extend | No | Yes | Unbiased | (Ameur et al. 2010) | http://solidsoftwaretools.com/gf/project/splitseek |
| Supersplat | Seed-and-extend | No | No | Unbiased | (Bryant et al. 2010) | https://github.com/mocklerlab/supersplat |
| SeqSaw | Seed-and-extend | No | Yes | Unbiased | (Wang et al. 2011) | http://bioinfo.au.tsinghua.edu.cn/software/seqsaw |
| ABMapper | Seed-and-extend | No | Yes | Exact match to GT-AG, GC-AG, AT-AC | (Lou et al. 2011) | http://abmapper.sourceforge.net/ |
| MapNext | Seed-and-extend | Optional | No | Known-junctions and GT-AG for novel ones | (Bao et al. 2009) | http://evolution.sysu.edu.cn/english/software/mapnext.htm |
| STAR | Seed-and-extend | Optional | Yes | Exact match to GT-AG, GC-AG, AT-AC and unbiased | (Dobin et al. 2012) | http://gingeraslab.cshl.edu/STAR/ |
| GSNAP | Seed-and-extend | No | Yes | Exact match to GT-AG, GC-AG, AT-AC | (Wu et al. 2010) | http://research-pub.gene.com/gmap/ |
| QPALMA | Seed-and-extend | No | No | SVM model for splice-sites | (De Bona et al. 2008) | http://www.raetschlab.org/suppl/qpalma |
| CRAC | Seed-and-extend | No | No | Unbiased | (Philippe et al. 2013) | http://crac.gforge.inria.fr/ |
| PALMapper | GenomeMapper + QPalma | No | Yes | Qpalma model | (Jean et al. 2010) | http://galaxy.raetschlab.org/ |
| CRAC | Multi-seed | No | No | Unbiased | (Philippe et al. 2013) | http://crac.gforge.inria.fr/ |
| OLEgo | Multi-seed | No | Yes | Combined model of splice-site sequence and intron length | (Wu et al. 2013) | http://zhanglab.c2b2.columbia.edu/index.php/OLego |
| Subread | Multi-seed | No | Yes | Exact match to GT-AG | (Liao et al. 2013) | http://bioconductor.org/packages/release/bioc/html/Rsubread.html |
| OSA | Seed-and-extend | Yes | Yes | Known and splice-sites and exact match to GT-AG, GC-AG, AT-AC | (Hu et al. 2012) | http://omicsoft.com/osa/ |
| X-MATE | Recursive mapping to genome and junctions | Yes | No | Known splice-sites | (Wood et al. 2011) | http://grimmond.imb.uq.edu.au/X-MATE/ |
| RNASEQR | Bowtie and BLAT on transcripts and genome | Yes | Yes | Known splice-sites and BLAT model | (Chen et al. 2012b) | https://github.com/rnaseqr/RNASEQR |
| MapAI | Bowtie alignments to transcripts | Yes | No | Known splice-sites | (Labaj et al. 2012) | http://www.bioinf.boku.ac.at/pub/MapAI |
| SAMMate | Bowtie to exons and junctions | Yes | Yes | Known splice-sites | (Xu et al. 2011) | http://sammate.sourceforge.net/ |
| IsoformEx | Bowtie to exons and junctions | Yes | No | Known splice-sites | (Kim et al. 2011) | http://bioinformatics.wistar.upenn.edu/isoformex |
| RUM | Bowtie and BLAT on transcripts and genome | Yes | Yes | Known splice-sites and BLAT model | (Grant et al. 2011) | http://www.cbil.upenn.edu/RUM/userguide.php |
| SpliceSeq | Bowtie alignments to Splicing graphs | Yes | Yes | Known splice-sites | (Ryan et al. 2012) | http://bioinformatics.mdanderson.org/main/SpliceSeq:Overview |
| PASTA | Bowtie alignment of read fragments | No | Yes | Logistic-regression model for splice-sites | (Tang et al. 2013) | http://genome.ufl.edu/rivalab/PASTA |
| ContextMap | Genome alignments from other methods | No | No | Unbiased | (Bonfert et al. 2012) | http://www.bio.ifi.lmu.de/softwareservices/contextmap |

**Table 2. Genome-based quantification of known events and isoforms.** This table includes methods that can be used to quantify known splicing events (RUM, SpliceSeq, MMES, SpliceTrap), known isoforms (SAMMate, IsoformEx, Erange, rSeq, rQuant, FluxCapacitor, IQSeq, Cufflinks, Casper, CEM, IsoInfer, SLIDE, RABT, DRUT, iReckon) or both (MISO, ALEXA-Seq, SOLAS) when a genome-based annotation is available. Some include the mapping step (RUM, SpliceSeq, SAMMate, IsoformEx). Some isoform-based methods can quantify known and novel isoforms simultaneously (IsoInfer, SLIDE, RABT, DRUT, iReckon), or choose between quantifying known or novel isoforms (Cufflinks, Casper, CEM, IsoLasso). We indicate the type of input used by the method in the cited reference, whether they exploit paired-end read information in the calculation and what type of quantification is given. We also provide the reference where the method is described, and the URL (or email) where the software is available.

| Method | Type | Input Used in publication | Uses paired-end reads | Quantification | Reference | Web site |
|---|---|---|---|---|---|---|
| RUM | Exon and junction quantification | Bowtie and BLAT on transcripts and genome | Yes | Read counts and RPKM of exons and junctions | (Grant et al. 2011) | http://www.cbil.upenn.edu/RUM/userguide.php |
| SpliceSeq | Exon and junction quantification | Bowtie alignments to Splicing graphs | Yes | Inclusion level of exons and junctions | (Ryan et al. 2012) | http://bioinformatics.mdanderson.org/main/SpliceSeq:Overview |
| MMES | Junction quantification | SOAP alignments to junctions | No | Junction scores | (Wang et al. 2010b) | Email to Wang.Liguo@mayo.edu |
| SpliceTrap | Exon and junction quantification | Bowtie on inclusion/skipping events | Yes (models insert sizes) | Exon inclusion level | (Wu et al. 2011) | http://rulai.cshl.edu/splicetrap/ |
| SAMMate | Isoform quantification | Bowtie on genome and junctions | Yes | RPKM/FPKM | (Xu et al. 2011) | http://sammate.sourceforge.net/ |
| IsoformEx | Isoform quantification | Bowtie on genome and junctions | No | Isoform expression (~RPKM) | (Kim et al. 2011) | http://bioinformatics.wistar.upenn.edu/isoformex |
| MISO | Event and isoform quantification | Bowtie on genome and junctions | Yes | Isoform PSI value | (Katz et al. 2010) | http://genes.mit.edu/burgelab/miso/ |
| ALEXA-Seq | Event and isoform quantification | Reads mapped to genome and junctions | Yes | Event and isoform expression level | (Griffith et al. 2010) | http://www.alexaplatform.org/alexa_seq/ |
| SOLAS | Event and isoform quantification | Reads mapped to genome | No | Isoform expression (~RPKM) | (Richard et al. 2010) | http://cmb.molgen.mpg.de/2ndGenerationSequencing/Solas/ |
| Erange | Isoform quantification | Bowtie on genome and junctions | No | Isoform RPKM | (Mortazavi et al. 2008) | http://woldlab.caltech.edu/rnaseq |
| rSeq | Isoform quantification | SeqMap alignments to exons and exon-exon junctions | Yes (in latest version) | Isoform RPKM | (Jiang et al. 2009) | http://www-personal.umich.edu/~jianghui/rseq/ |
| rQuant | Isoform quantification | Reads mapped to genome | No | Isoform average read coverage and RPKM | (Bohnert et al. 2009) | http://galaxy.raetschlab.org/ |
| FluxCapacitor | Isoform quantification | Reads mapped to genome | Yes | Isoform relative abundance (~PSI) | (Montgomery et al. 2010) | http://flux.sammeth.net/capacitor.html |
| IQSeq | Isoform quantification | GFF/MRF/BED | Yes | Isoform RPKM | (Du et al. 2012) | http://archive.gersteinlab.org/proj/rnaseq/IQSeq/ |
| Cufflinks | Known or novel isoform quantification | TopHat alignments | Yes | FPKM | (Trapnell et al. 2010) | http://cufflinks.cbcb.umd.edu/ |
| Casper | Known or novel isoform quantification | TopHat alignments | Yes | Isoform PSI value | (Rossell et al. 2012) | https://sites.google.com/site/rosselldavid/software |
| CEM | Known or novel isoform quantification | TopHat alignments | Yes | Isoform expression | (Li et al. 2012a) | http://alumni.cs.ucr.edu/~liw/cem.html |
| IsoLasso | Known or novel isoform quantification | TopHat alignments | Yes | RPKM | (Li et al. 2011a) | http://alumni.cs.ucr.edu/~liw/isolasso.html |
| IsoInfer | Known and novel isoform quantification | TopHat alignments | Yes | Isoform RPKM | (Feng et al. 2012) | http://www.cs.ucr.edu/~jianxing/IsoInfer.html |
| SLIDE | Known and novel isoform quantification | modEncode spliced mappings | Yes | Isoform RPKM | (Li et al. 2011b) | https://sites.google.com/site/jingyijli/SLIDE.zip |
| RABT | Known and novel isoform quantification | TopHat alignments | Yes | Isoform FPKM | (Roberts et al. 2011) | http://cufflinks.cbcb.umd.edu/ |
| DRUT | Known and novel isoform quantification | Bowtie/TopHat alignments to transcriptome/genome | No | FPKM (computed by IsoEM) | (Mangul et al. 2012) | http://www.cs.gsu.edu/~serghei/?q=drut |
| iReckon | Known and novel isoform quantification | TopHat alignments | Yes | Isoform RPKM | (Mezlini et al. 2013) | http://compbio.cs.toronto.edu/ireckon/ |

**Table 3. Isoform quantification guided by a transcriptome.** This table includes methods that quantify isoforms using a transcriptome annotation and reads mapped with a non spliced-mapper. All the methods used bowtie to map reads to transcripts in the original publication. Although they generally work with reads mapped to a transcriptome, some methods (RSEM, MMSEQ). We indicate the type of input used by the method, whether they exploit paired-end read information in the calculation and what type of isoform quantification is given. We also provide the reference where the method is described, and the URL where the software is available.

| Method | Input reads format | Uses paired-end reads | Isoform Quantification | Reference | Web site |
|---|---|---|---|---|---|
| RSEM | BAM/SAM | Yes (models insert size) | "Expected number of fragments per isoform" and "fraction of transcripts represented by the isoform" | (Li et al. 2011c) | https://github.com/bli25wisc/RSEM/ |
| IsoEM | SAM | Yes (models insert size) | Isoform expression | (Nicolae et al. 2011) | http://dna.engr.uconn.edu/?page_id=105 |
| NEUMA | Fastq/Fasta mapped with Bowtie | Yes | FVKM (fragments per virtual kilobase per million sequenced reads) | (Lee et al. 2011) | http://neuma.kobic.re.kr |
| BitSeq | SAM | Yes (models insert size) | Isoform expression | (Glaus et al. 2012) | http://www.bioconductor.org/packages/2.11/bioc/html/BitSeq.html |
| MMSEQ | Sorted BAM | Yes | Haplotype and isoform-specific expression | (Turro et al. 2011) | http://bgx.org.uk/software/mmseq.html |
| eXpress | BAM | Yes | FPKM, estimated counts | (Roberts et al. 2013) | http://bio.math.berkeley.edu/eXpress/ |

**Table 4. Genome-based reconstruction and quantification without annotation.** This table includes methods to reconstruct (all methods) and to quantify (all methods except for G-Mo.R-Se and assemblySAM) multiple isoforms from genome-mapped reads without using any gene annotation. Some methods can also be run with annotations for quantification (Cufflinks, IsoLasso, Casper, CEM). Some perform simultaneously reconstruction and quantification of novel isoforms (NSMAP, Montebello, IsoLasso). We indicate the type of input used by the method in the cited reference, whether they exploit paired-end read information in the calculation and what type of isoform quantification is given. We also provide the reference where the method is described, and the URL or email where the software is available.

| Method | Type | Input used in publication | Uses paired-end reads | Isoform Quantification | Reference | Web site |
|---|---|---|---|---|---|---|
| G-Mo.R-Se | *De novo* isoform reconstruction | SOAP alignments | No | No | (Denoeud et al. 2008) | http://www.genoscope.cns.fr/externe/gmorse/ |
| assemblySAM | *De novo* isoform reconstruction | Own heuristics for read-mapping using Bowtie | Yes | No | (Zhao et al. 2011a) | http://sammate.sourceforge.net/assemblysam.html |
| TAU | *De novo* isoform reconstruction and quantification | Supersplat alignments | No | Average per-base sequencing depth | (Filichkin et al. 2010) | Email to HPriest@danforthcenter.org |
| Scripture | *De novo* isoform reconstruction and quantification | TopHat alignments | Yes (models insert size) | RPKM | (Guttman et al. 2010) | http://www.broadinstitute.org/software/Scripture/ |
| Cufflinks | Known or novel isoform quantification | TopHat alignments | Yes (models insert size) | FPKM | (Trapnell et al. 2010) | http://cufflinks.cbcb.umd.edu/ |
| Casper | Known or novel isoform quantification | TopHat alignments | Yes | Isoform PSI value | (Rossell et al. 2012) | https://sites.google.com/site/rosselldavid/software |
| CEM | Known or novel isoform quantification | TopHat alignments | Yes | Isoform expression | (Li et al. 2012a) | http://alumni.cs.ucr.edu/~liw/cem.html |
| IsoLasso | Known or novel isoform quantification | TopHat alignments | Yes | RPKM | (Li et al. 2011a) | http://alumni.cs.ucr.edu/~liw/isolasso.html |
| Montebello | Novel isoform reconstruction and quantification | SpliceMap alignments | Yes | Isoform expression | (Hiller et al. 2012) | http://www.stanford.edu/group/wonglab/Montebello/Montebello_0.8.tar.gz |
| NSMAP | Novel isoform reconstruction and quantification | TopHat alignments | Yes (models insert size) | RPKM | (Xia et al. 2011) | https://sites.google.com/site/nsmapforrnaseq |

**Table 5. Evidence-based alternatively spliced gene prediction.** This table includes methods that could be used to perform alternatively spliced gene prediction from RNA-Seq data. Besides the *de novo* reconstruction and quantification methods from Table 4 and those methods from Table 2 that can predict novel and known isoforms simultaneously (IsoInfer, SLIDE, RABT, DRUT, iReckon), there are also methods that can use various sources of evidence to predict alternatively spliced genes (TAU, SpliceGrapher, ExonMap/JunctionWalk) and methods that predict alternatively spliced protein coding genes from multiple evidences (Augustus, mGene). We also include classical protein-coding gene prediction methods that could potentially use RNA-Seq as evidence (Gaze, JigSaw, EVM, Evigan). For each method, we indicate the type of input used, whether they exploit paired-end read information in the calculation or provide any isoform quantification. We also give the reference where the method is described and the URL or email where the software is available.

| Method | Type | Input Used in publication | Uses paired-end reads | Isoform quantification | Reference | Web site |
|---|---|---|---|---|---|---|
| IsoInfer | Known and novel isoform quantification | TopHat alignments | Yes | Isoform RPKM | (Feng et al. 2012) | http://www.cs.ucr.edu/~jianxing/IsoInfer.html |
| SLIDE | Known and novel isoform quantification | modEncode spliced mappings | Yes | Isoform RPKM | (Li et al. 2011b) | https://sites.google.com/site/jingyijli/SLIDE.zip |
| RABT | Known and novel isoform quantification | TopHat alignments | Yes | Isoform FPKM | (Roberts et al. 2011) | http://cufflinks.cbcb.umd.edu/ |
| DRUT | Known and novel isoform quantification | Bowtie (TopHat) alignments to transcriptome (genome) | No | FPKM (computed by IsoEM) | (Mangul et al. 2012) | http://www.cs.gsu.edu/~serghei/?q=drut |
| iReckon | Known and novel isoform quantification | TopHat alignments | Yes | Isoform RPKM | (Mezlini et al. 2013) | http://compbio.cs.toronto.edu/ireckon/ |
| TAU | Evidence-based isoform reconstruction and quantification | Supersplat alignments | No | Average per-base sequencing depth | (Filichkin et al. 2010) | Email to hpriest@danforthcenter.org |
| SpliceGrapher | Evidence-based isoform reconstruction | TopHat alignments | Yes | No | (Rogers et al. 2012) | http://SpliceGrapher.sf.net |
| ExonMap/JunctionWalk | Evidence-based isoform reconstruction | Reads mapped to exons and junctions | Handled by SpliceMap | No | (Seok et al. 2012a) | http://gluegrant1.stanford.edu/~DIC/RNASeqArray/TranscriptConstruction.html |
| mGene | Evidence-based alternatively-spliced gene prediction | Reads mapped to genome | Yes | No | (Behr et al. 2010) | http://mgene.org/ |
| Augustus | Evidence-based alternatively-spliced gene prediction | Spliced evidences | No | No | (Stanke et al. 2006a) | http://bioinf.uni-greifswald.de/augustus/ |
| Gaze | Evidence-based gene prediction | Evidence in GFF format | No | No | (Howe et al. 2002) | http://www.sanger.ac.uk/resources/software/gaze/ |
| JigSaw | Evidence-based gene prediction | Spliced evidences | No | No | (Allen et al. 2005) | http://www.cbcb.umd.edu/software/jigsaw/ |
| EVM | Evidence-based gene prediction | PASA alignments | No | No | (Haas et al. 2008) | http://evidencemodeler.sourceforge.net/ |
| Evigan | Evidence-based gene prediction | Spliced evidences | No | No | (Liu et al. 2008) | http://www.seas.upenn.edu/~strctlrn/evigan/evigan.html |

**Table 6. *De novo* transcriptome assembly.** This table includes methods for *de novo* transcriptome assembly. Some of these methods produce multiple isoforms per assembled gene (OASES, SOAPdenovo-trans, TransAbyss, Trinity), and only two quantify the alternative isoforms (TransAbyss, Trinity). Nonetheless, these methods can in theory be coupled with transcriptome-based quantification methods (Table 3). KisSplice assembles alternatively spliced events rather than isoforms and quantifies the read coverage of these events. We indicate whether they exploit paired-end read information in the calculation, use a single/multiple k-mer approach, detect multiple isoforms per gene or quantify isoforms. We also provide the reference where the method is described and the URL (or email) where the software is available.

| Method | Uses paired-end reads | Graph approach | Detects alternative isoforms | Isoform quantification | Reference | Web site |
|---|---|---|---|---|---|---|
| Rnnotator | Yes | Variable k-mer | No | No | (Martin et al. 2010) | Email to vtdelapuente@lbl.gov |
| STM | Yes | Variable k-mer | No | No | (Surget-Groba et al. 2010) | http://www.surget-groba.ch/downloads/stm.tar.gz |
| OASES | Yes | Variable k-mer | Yes | No | (Schulz et al. 2012) | http://www.ebi.ac.uk/~zerbino/oases/ |
| SOAPdenovo-trans | Yes | Variable k-mer | Yes | No | (Li et al. 2009) | http://soap.genomics.org.cn/SOAPdenovo-Trans.html |
| TransAbyss | Yes | Variable k-mer | Yes | Isoform read coverage | (Robertson et al. 2010) | http://www.bcgsc.ca/platform/bioinfo/software/ |
| Trinity | Yes | Single k-mer | Yes | Yes (uses RSEM) | (Grabherr et al. 2011) | http://TrinityRNASeq.sourceforge.net |
| KisSplice | No | Single k-mer | Events only | Event read coverage | (Sacomoto et al. 2012) | http://alcovna.genouest.org/kissplice/ |

**Table 7. Differential splicing.** These methods measure changes in inclusion between two or more conditions at the exon level (DEXSeq, DSGSeq, GPSeq, SOLAS), event level (MATS, JuncBASE, JETTA, SpliceSeq), and isoform region level (DiffSplice, SplicingCompass, FDM, rDiff) or at both, isoform and event inclusion (MISO, ALEXA-Seq), We indicate whether the methods perform any quantification per sample, whether they exploit paired-end read information in the calculation, what is the measure of differential splicing provided, the reference where the method is described, and the URL where the software is available.

| Method | Type | Quantification | Uses paired-end reads | Models biological variability | Differential quantification | Reference | Web site |
|---|---|---|---|---|---|---|---|
| DEXSeq | Exon level | No | No | Yes | Differential exon inclusion | (Anders et al. 2012) | http://www.bioconductor.org/packages/release/bioc/html/DEXSeq.html |
| DSGSeq | Exon level | Isoform relative abundances | No | Yes | Differential exon inclusion | (Wang et al. 2013) | http://bioinfo.au.tsinghua.edu.cn/software/DSGseq |
| GPSeq | Exon level | Exon splicing ratio | No | Yes | Differential exon splicing index | (Srivastava et al. 2010) | http://cran.r-project.org/web/packages/GPseq/index.html |
| SOLAS | Exon level | Event and isoform inclusion | No | No | Differential exon inclusion | (Richard et al. 2010) | http://cmb.molgen.mpg.de/2ndGenerationSequencing/Solas/ |
| MATS | Event level | Event inclusion | Handled my mapping method | Yes | Differential event inclusion | (Shen et al. 2012) | http://rnaseq-mats.sourceforge.net/ |
| JuncBASE | Event level | Event inclusion | Yes | No | Differential event inclusion | (Brooks et al. 2011) | http://compbio.berkeley.edu/proj/juncbase/Home.html |
| JETTA | Event level | SeqMap alignments | Handled by mapping method | No | Differential event inclusion | (Seok et al. 2012b) | http://igenomed.stanford.edu/~junhee/JETTA/rnaseq.html |
| SpliceSeq | Event level | Inclusion level of exons and junctions | Yes | No | Differential event inclusion | (Ryan et al. 2012) | http://bioinformatics.mdanderson.org/main/SpliceSeq:Overview |
| Alexa-Seq | Event and isoform levels | Gene, transcript, and event expression levels | Yes | No | Differential relative event/isoform expression | (Griffith et al. 2010) | http://www.alexaplatform.org/alexa_seq/ |
| MISO | Event and isoform levels | PSI | Yes | No | Differential event/isoform PSI | (Katz et al. 2010) | http://genes.mit.edu/burgelab/miso/ |
| SplicingCompass | Isoform-region-level | Normalized exon density | Handled my mapping method | No | Differential relative isoform abundance | (Aschoff et al. 2013) | http://www.ichip.de/software/SplicingCompass.html |
| DiffSplice | Isoform region level | Expression of "Alternative Splicing Modules" | Yes | Yes | Differential Expression of "Alternative Splicing Modules" | (Hu et al. 2013) | http://www.netlab.uky.edu/p/bioinfo/DiffSplice |
| FDM | Isoform region level | Isoform region relative expression | No | Yes | Differential relative isoform abundance | (Singh et al. 2012) | http://csbio-linux001.cs.unc.edu/nextgen/software/FDM |
| rDiff | Isoform region level | Isoform region relative expression | Yes | Yes | Differential relative isoform abundance | (Drewe et al. 2013) | http://bioweb.me/rdiff |

**Table 8. Isoform-based differential expression.** These methods measure differential expression at the transcript level between two or more conditions, allowing multiple transcripts per gene. Cuffdiff2 additionally can calculate significant changes in the relative abundance of isoforms. For each method, we indicate the quantification performed per sample, whether it exploits paired-end read information in the calculation, the measure of differential expression provided, the reference where the method is described and the URL where the software is available.

| Method | Quantification | Uses paired-end reads | Models biological variability | Differential quantification | Reference | Web site |
|---|---|---|---|---|---|---|
| BASIS | Isoform relative expression | No | No | Differential isoform expression | (Zheng et al. 2009) | http://www-rcf.usc.edu/~liangche/software.html |
| Cuffdiff2 | Isoform expression | Yes | Yes | Differential isoform expression | (Trapnell et al. 2012) | http://cufflinks.cbcb.umd.edu/ |
| BitSeq | Isoform expression | Yes | Yes | Differential isoform expression | (Glaus et al. 2012) | http://www.bioconductor.org/packages/2.11/bioc/html/BitSeq.html |
| EBSeq | Isoform expression quantified by input method | Handled by quantification method | Yes | Differential isoform expression | (Leng et al. 2013) | http://www.biostat.wisc.edu/~kendzior/EBSEQ/ |

**Table 9. Visualizing Alternative Splicing.** This table includes some of the available tools for the visualization of alternative splicing using RNA-Seq data. Some of them can be used as command line tools that are included in the distribution of the analysis tools (RSEM, SpliceGrapher, DiffSplice, DEXSeq, SplicingCompass) or provided separately (Sashimi Plots), whereas others are Graphical User Interfaces (Savant, ALEXA-Seq, SpliceSeq).

| Method | Type | Used with | Input data | Visualization | Reference | Web site |
|---|---|---|---|---|---|---|
| RSEM | Command line tool | RSEM | Transcript BAM file | Read profiles (WIG) | (Li et al. 2011c) | https://github.com/bli25wisc/RSEM/ |
| SpliceGrapher | Command line tool | SpliceGrapher | GFF files | Isoforms | (Rogers et al. 2012) | http://SpliceGrapher.sf.net |
| DiffSplice | Command line tool | DiffSplice | GTF (graphs) | Isoforms | (Hu et al. 2012) | http://www.netlab.uky.edu/p/bioinfo/DiffSplice |
| DEXSeq | Command line tool | DEXSeq | DEXSeq results | Differential exon usage | (Anders et al. 2012) | http://www.bioconductor.org/packages/release/bioc/html/DEXSeq.html |
| SplicingCompass | Command line tool | SplicingCompass | SplicingCompass results | Differential exon usage | (Aschoff et al. 2013) | http://www.ichip.de/software/SplicingCompass.html |
| Sashimi Plots | Command line tool | MISO | GFF3 | Splicing events and read coverage | (Katz et al. 2010) | http://genes.mit.edu/burgelab/miso/docs/sashimi.html |
| Savant Browser | GUI | iReckon | GFF | Isoforms | (Fiume et al. 2010) | http://genomesavant.com/savant/ |
| ALEXA-Seq viewer | GUI | ALEXA-Seq | Alexa-seq database | Splicing events and expression levels | (Griffith et al. 2010) | http://www.alexaplatform.org/alexa_seq/ |
| SpliceSeq | GUI | SpliceSeq | SpliceSeq processed data | Isoforms and alternatively spliced events | (Ryan et al. 2012) | http://bioinformatics.mdanderson.org/main/SpliceSeq:Overview |